\documentclass[twocolumn,epsfig,color]{mn2e}
\usepackage{hyperref}
\usepackage[usenames]{color}
\hypersetup{dvips, colorlinks=true, linkcolor=blue, citecolor=blue, filecolor=blue, urlcolor=blue}
\usepackage[dvips]{graphicx}
\usepackage{amssymb}

\newcommand{\mincir}{\raise
-2.truept\hbox{\rlap{\hbox{$\sim$}}\raise5.truept
\hbox{$<$}\ }}
\newcommand{\magcir}{\raise
-2.truept\hbox{\rlap{\hbox{$\sim$}}\raise5.truept
\hbox{$>$}\ }}
\newcommand{\minmag}{\raise-2.truept\hbox{\rlap{\hbox{$<$}}\raise
6.truept\hbox{$>$}\ }}
\newcommand{\mnras}{MNRAS}
\newcommand{\apj}{ApJ}
\newcommand{\apjl}{ApJL}
\newcommand{\apjs}{ApJS}

\newcommand{\aap}{A\&A}
\newcommand{\araa}{ARA\&A}
\newcommand{\nat}{Nature}

\newcommand{\Sim}{\mbox{{\sc \small Sim}}}
\newcommand{\Halo}{\mbox{{\sc \small Halo}}}

\title[Evolution of the distribution of baryons in a simulated Local Group]
{ Evolution of the  baryon fraction in the Local Group: \\
  accretion versus feedback at low and high z}

\author[S.~Peirani et al.]
 {S\'ebastien~Peirani$^{1}$\thanks{E-mail: peirani@iap.fr},
  Intae Jung$^{1,2}$,
  Joseph~Silk$^{1,3}$ and Christophe Pichon$^{1,4}$\\
$^{1}$ Institut d'Astrophysique de Paris (UMR 7095: CNRS \& UPMC), 98 bis Bd Arago, 75014 Paris, France \\
$^{2}$ Department of Astronomy, Yonsei University, Seoul 120-749, Korea\\
$^{3}$ Department of Physics, University of Oxford, Denys Wilkinson Building,
 Keble Road, Oxford OX1 3RH, UK\\
$^{4}$ Institut de Physique Th\'eorique, Orme des Merisiers, batiment 774, CEA/Saclay
F-91191 Gif-sur-Yvette, France.\\
}

\begin{document}

\maketitle

\begin{abstract}

Using hydrodynamical zoom simulations in the standard $\Lambda$CDM cosmology, we
investigate the evolution of the distribution of baryons (gas and stars) in a 
local group-type universe. \\
First, with standard star formation and supernova feedback
prescriptions,  we find that
 the mean baryonic fraction value estimated at the virial radius
of the two main central objects (i.e. the Milky Way and Andromeda) is decreasing over
time, and  is  10-15\% lower than the universal value, 0.166, at $z=0$.
This decrease is mainly due to
 the fact that the amount of  accretion of  dissipative gas onto the halo,
especially at low redshift, is in general much lower than that  of the dissipationless dark matter.
Indeed,
a significant part of the baryons does not collapse onto the haloes and remains in their outskirts, mainly in the form
of warm-hot intergalactic medium (WHIM).  
Moreover, during the formation of each object, some  dark matter and baryons are also expelled 
through  merger events via tidal disruption.  
In contrast  to baryons,  expelled dark matter  can be more efficiently 
 re-accreted  onto the halo, enhancing both the reduction of $f_b$ inside $R_v$, 
and the increase of the
mass of WHIM outside $R_v$.
Varying  the efficiency of supernovae feedback at low redshift does not seem to  significantly affect these trends.
 \\
 Alternatively, when a significant fraction of the initial gas in the main objects
is released at high redshifts by more powerful sources of feedback, 
such as AGN from intermediate mass black holes in lower mass galaxies, 
the baryonic fraction at the virial radius can have a lower value ($f_b\sim 0.12$) at low redshift. 
Hence physical mechanisms able to drive the gas out of the virial radius at high redshifts will 
have a stronger impact on the deficit of baryons in the mass budget 
of Milky Way type-galaxies at present times 
 than those that expel the gas  in the longer,  late phases of galaxy formation.

\end{abstract}

\begin{keywords}
Galaxies: Local Group -- Galaxies: haloes -- Dark matter --  Methods: N-body simulations 
\end{keywords}

\section{Introduction}

The distribution of galaxies in the Universe on large scales, as revealed by sky surveys
such as the CFA Redshift Survey (Geller \& Huchra 1989), 
 the 2DF Galaxy Redshift Survey (Colless et al. 2001)
or the Sloan Digital Sky Survey (Gott et al. 2005)
is structured into a filamentary web in which the intersections
{correspond } to  the most massive objects (i.e. galaxy clusters, Bond et al. 1996).
This specific distribution is
now commonly  and easily reproduced by cosmological hydrodynamical 
N-body simulations of the $\Lambda$CDM cosmology which also indicate 
that the distribution of dark matter, the dominant mass component of the Universe (Komatsu et al. 2011),
follows the same trend. Indeed,  since in the traditional picture of galaxy formation, galaxies are supposed to
form when baryonic gas falls into the gravitational potential of their host dark matter halo, the distribution
of dark matter is therefore expected to faithfully trace that of the baryons.
However on galactic scales, observations tend to suggest that the spatial distributions
of dark matter and baryons (especially in the form of gas) may display
some substantial differences. In particular, it has been shown that galaxies are missing
most of their baryons, --  most galaxies are severely baryons-depleted relative to
the cosmological fraction (see for instance Bell et al. 2003; Hoekstra et al. 2005; McGaugh 2010). 

This so-called ``missing baryons problem" (see Bregman 2007 for a complete review),
if {\sl real}, calls for two alternative scenarios. Either a significant part of the gas never collapsed
into the gravitational potential wells of protogalaxies in the first place,
or some of the gas has been expelled by  galaxy formation  feedback processes such as 
supernova winds. Hence solving the 
 missing baryon problem may prove to be central in order to  constrain  galaxy formation models.

Past studies based on hydrodynamical simulations indicate that most of
the ``missing'' baryons might lie in a gaseous phase (the so-called warm-hot intergalactic medium, hereafter WHIM)
in the temperature range $10^5-10^7$ $K$  and at moderate over-density (Cen \& Ostriker 1999; Dav\'e et
al. 2001; Cen \& Ostriker 2006). According to these studies, the WHIM is primarily shock-heated during the formation of
large-scale structures, while feedback mechanisms associated with star formation
should also have an additional impact on this phase during galaxy formation.
For instance, for low mass systems, the cosmological UV-background is supposed to reduce both 
the star formation and baryon content (Thoul \& Weinberg 1996; Bullock, Kravtsov \& Weinberg 2000; 
Gnedin 2000; Somerville 2002; Ricotti, Gnedin \& Shull 2002;
Benson et al. 2002; Read, Pontzen \& Viel 2006; 
 Hoeft et al. 2006;  Okamoto, Gao \& Theuns 2008; Peirani 2010;
Nickerson et al. 2011, 	Hambrick et al. 2011a).  Supernova feedback is also expected to 
expel some gas, especially for low mass systems, as suggested by various numerical investigations
(Scannapieco et al. 2008, 2009; Faucher-Gigu{\`e}re, Kere{\v s} \& Ma 2011 and references therein).
Other mechanisms have been proposed, such as  turbulence in the baryonic intergalactic medium  (Zhu, Feng \& Fang 2011)
or pre-heating by pre-virialisation (Mo et al. 2005), although Crain et al. (2007) claimed that this process is
unable to prevent the collapse of gas  
 (instead, according to these authors, non-gravitational feedback is required in order to reduce the efficiency of gas cooling and star formation in dwarf 
galaxies).
Another less investigated potential mechanism, which may also affect the evolution 
of the distribution of baryons on small scales,
is the effect of accretion and merger events (and
 in particular their associated tidal disruption) that are expected to be frequent in the framework of the hierarchical model.
For instance, in an early paper, Navarro \& White (1993)
have shown that during a merger involving dark matter and adiabatic
gas, there is a transfer of energy between these two components that leads to a situation where the gas is less tightly bound.
More recently, Sinha \& Holley-Bockelmann (2010, 2011) found that a few percent of gas can be driven into the 
intergalactic medium (IGM) by galaxy mergers,
 using either merger trees calculated from the Press-Schechter formalism
or idealized simulations of mergers of galaxy clusters. Using hydrodynamical simulations, Nickerson et al. (2011)
have also concluded that tidal forces may cause significant mass loss from satellites of all masses.

In the present work, we make  use of cosmological ``zoom'' simulations with an extended
treatment of the physics of baryons
to study the formation of Milky Way-like galaxies. Our aim is to characterise the  relative
 role of supernova feedback   to accretion and mergers 
 in the evolution of the distribution of baryons for objects of such masses. We will also
test scenarios in which  a significant fraction of gas in progenitors 
is expelled at high redshift by
 more powerful sources of feedback,  such as AGN associated  with massive black holes. 
This  will allow us to quantify  two distinct processes which may allow us to  { address} the so-called
 missing baryon problem, should it persist.
 
This paper is organised as follows. In section 2,  we summarize the numerical modelling, and section 3 presents our main results
on the evolution of the distribution of baryons in our simulated local group universes. 
We give our main conclusions in the last section.

\section{Numerical modelling}

\subsection{Initial conditions and simulation }

The numerical methodology used in the present paper is described
in detail in Peirani (2010)  to which we refer the reader for more information.
For the sake of clarity, we summarize the main steps below.

We analyse three cosmological zoom simulations  for  a   $\Lambda$CDM
universe  using WMAP5 parameters (Komatsu et al. 2009), namely
$\Omega_{\rm M}=0.274$,  $\Omega_{\Lambda}=0.726$, $\Omega_b=0.0456$,  $H_0=70.5$
km/s/Mpc, $n=0.96$ and $\sigma_8=0.812$. 
Each simulation was performed in a periodic box of side $100\,h^{-1}$ Mpc
with $2\times 2048^3$ effective dark matter and gas particles in the highest resolution region (a sphere of
$7\,h^{-1}$ Mpc of radius). In this region of interest, the  mass
resolution are $m_{\rm DM}\approx 7.4\times 10^6\,h^{-1} M_\odot$ and
$m_{\rm gas}=m_{\rm stars}\approx 1.5\times 10^6\,h^{-1} M_\odot$.
The Plummer-equivalent force softening adopted for the
high mass resolution particles were 1 and 0.5 $h^{-1}$ kpc
for dark matter and gas particles respectively
 and were kept constant in comoving units.

Initial conditions have been generated from the MPgrafic code (Prunet et
al. 2008) and the simulations were performed using GADGET2 
(Springel 2005), with added  prescriptions for metal-dependent cooling, star
formation (in this work we remind that the star formation efficiency is $c_*$=0.02),
 feedback from Type Ia and II supernovae (SN), UV background (starting at $z=8.5$)
and metal enrichment.
The three simulations have common initial conditions whose phases are consistent with the local group
 but essentially differ in  the quantity of energy released
by SN derived from star particles $i$ ($E_i$).
 As mentioned in Peirani (2010), we consider that a fraction $\gamma$ of
this energy is deposited in the $j^{th}$ neighbour gas particle by
applying a radial kick to its velocity with a magnitude $\Delta
v_j = \sqrt{(2w_j\gamma E_i/m_j)}$, where $w_j$ is the weighting
based on the smoothing kernel and $m_j$ is the mass of gas
particle $j$. The first simulation $\Sim1a$ uses the standard value of $\gamma=0.1$ while in the second one, $\Sim1b$,
we have considered a higher efficiency $\gamma=1.0$ in order to investigate how our results would be affected.
In the third simulation ($\Sim2$), our aim is to study the effects of earlier high energy ejection
to the ISM induced either by  intermediate mass 
black holes or other high energy processes such as hypernovae events.
 Intermediate mass black holes are the likely missing link between
 Population III and the supermassive black holes in quasars   and plausibly
 as part of the hierarchy of structure formation.  
  Hypernovae are likely to be more prevalent than normal supernovae in the earliest phases
 of star formation: for example, the Population III IMF is most likely top-heavy, and this may also be 
the norm for  the precursors of spheroids, as evidenced for example by the radial
 distribution of SNe in disturbed (including recently merged)  galaxies (Habergham,  Anderson \&  James 2010)
and the frequency of low mass x-ray binaries in ultracompact dwarf galaxies
(Dabringhausen et al. 2012).
A higher frequency of hypernovae could plausibly provide an order-of-magnitude higher
feedback efficiency than supernovae. The case of AGN is intriguing, as even higher
feedback is required in deep potential wells, where supernovae are relatively inefficient,
in order to account for the correlation between black hole mass and spheroid velocity
dispersion (Silk \& Nusser 2010; Debuhr et al. 2012). 
Moreover, the recent detection of a massive gas outflow in a distant quasar ($z=6.4$) strongly
suggests that  a strong quasar activity is already in place at very early times (Maiolino et al. 2012).  
For this purpose, a simple modelling was used in which
a much higher efficiency ($\gamma=50$) was considered 
during a very short ($\Delta t\sim 45$\,Myr) at earlier times ($z \sim 8.0$)
 and $\gamma=0.1$ otherwise. 
We also justify the choice of this high efficiency by 
simple energetics and momentum comparisons between supernovae and AGN.
For instance, for a $10^7$\,M$_\odot$ black hole along with its $10^{10}$\,M$_\odot$ in stars, the
number of type II SNe  produced in star formation phase ($\sim 150$\,M$_\odot$  per SN for a Chabrier IMF) is 
$\sim 7.10^7$. Thus, the total SNe energy injected is  $\sim 10^{59}$\,ergs over a dynamical time,
say $t_{\rm dyn}\sim 10^8$\,yr, 
or $10^{43.5}$\,ergs/s. Since the Eddington luminosity is $L_{Eddington}\sim 10^{45}$\,ergs/s energetics favour AGN.
Furthermore, for supernovae momentum conservation starts at shell velocity $\sim 400$\,km/s.
Since the ejection velocity is $\sim 2.10^4$\,km/s, the momentum injected is $\sim 2\%$ of initial momentum
namely $\sim 10^{48}$\,gm\,cm/s or $\sim 10^{32.5}$\,dynes.
For AGN, momentum injected is $L_{Eddington}/c$ or $\sim10^{34.5}$\,dynes. Thus AGN injects 100 times more momentum flux
than supernovae. Finally, there is another factor of 10-100 that favours AGN, the so-called mechanical advantage factor due 
to ram pressure on the expanding bubble as suggested by recent hydrodynamical simulations
(Wagner, Bicknell \& Umemura). Therefore the value of $\gamma=50$ seems to be reasonable in order
to account for the accumulation of the different potential sources of feedback at high redshifts. 

The feedback  parameters are summarized in Table \ref{table1}.
 
\begin{table}
\begin{center}
\caption{Fractions $\gamma$ of the energy released by supernovae
 } 
\begin{tabular}{llll}
\hline
&  $\Sim1a$  & $\Sim1b$ & $\Sim2$   \\
\hline
$\gamma$&  0.1  & 1.0  &  50 ($8.1\geq z \geq 7.7$)  \quad   0.1 elsewhere  \\
\end{tabular}
\label{table1}
\end{center}
\end{table}

\subsection{Physical properties of  main galaxies}

In each simulation, we analyse a  pair of galaxies with physical
characteristics similar to the Milky Way-Andromeda pair (MW-M31) and
with similar galaxy environments up to $7\,h^{-1}$ Mpc. Some
relevant physical properties at the virial radius $R_v$ at $z=0$ of these two objects from
each simulation are summarized in Table \ref{table2}.  Note that we keep the same definition of
the virial radius $R_V$ and virial mass $M_V$ used in Peirani (2010) (see paragraph 2.3), namely
$R_V$ is the radius where the enclosed mean density $M_V/(4\pi R_V^3/3)$
is $\Delta_c$ times the critical density, and
$\rho_c(z)=3H(z)^2/8\pi G$, where $H(z)=H_0\sqrt{\Omega_{\rm M}(1+z)^3+\Omega_\Lambda}$.
In the cosmology adopted in the present study, $\Delta_c=97.6$ at $z=0$.
Furthermore, the baryonic fraction is defined as follow:
%
\begin{equation}
 f_b\equiv\frac{m_{\rm baryons}}{m_{\rm baryons}+m_{\rm DM}},
\end{equation}
where 
$m_{\rm baryons}$ and $m_{\rm DM}$ refer to the masses of baryons (gas+stars) and dark matter respectively in the same specific 
region of the universe. When not specified in the text, we do not distinguish 
the cold and hot phases of the gas in the estimate of $f_b$.
According to the adopted cosmology, the universal  $f_b$ value is:
\begin{equation}
\langle f_b \rangle \equiv \frac{\Omega_b}{\Omega_{\rm M}} \approx 0.166.
\end{equation}

\begin{table*}
\begin{center}
\caption{Physical properties of the 2 main central objects of each simulation derived at the virial radius $R_{v}$ and $z=0$. } 
\begin{tabular}{cccccccc}
\hline
&   & $\Halo1$  &  &  &&  $\Halo2$ &   \\
\hline
&  $\Sim1a$  & $\Sim1b$  & $\Sim2$ && $\Sim1a$ & $\Sim1b$ & $\Sim2$  \\
& ($\gamma=0.1$)  & ($\gamma=1.0$)  &  && ($\gamma=0.1$) & ($\gamma=1.0$) &   \\
\hline
$m_{\rm DM}  \; (\times 10^{11}  h^{-1}M_{\odot}) $ & 7.58  & 7.58  &7.23&    & 7.43 &7.40  &  8.17   \\

$m_{\rm gas}  \; (\times 10^{11} h^{-1}M_{\odot}) $ &  0.30& 0.44&0.27&  & 0.38 &0.44  & 0.45\\

$m_{\rm stars}  \; (\times 10^{11} h^{-1}M_{\odot}) $ &  1.08 &0.93&0.71   &&0.84&0.78 & 0.68\\

$M_{v}  \; (\times 10^{11} h^{-1}M_{\odot})   $ &  8.96 &8.95&8.21   &&8.65&8.62 &9.30  \\

$R_{v} \; ({\rm Mpc})$ & 0.282  &0.282& 0.274  & &0.279 &0.279& 0.286 \\

$f_{b}$ &  0.154  &0.153 &0.119 && 0.141 &0.141 &0.121 \\
\hline
\end{tabular}
\label{table2}
\end{center}
\end{table*}

\begin{figure}
\begin{center}
\rotatebox{0}{\includegraphics[width=\columnwidth]{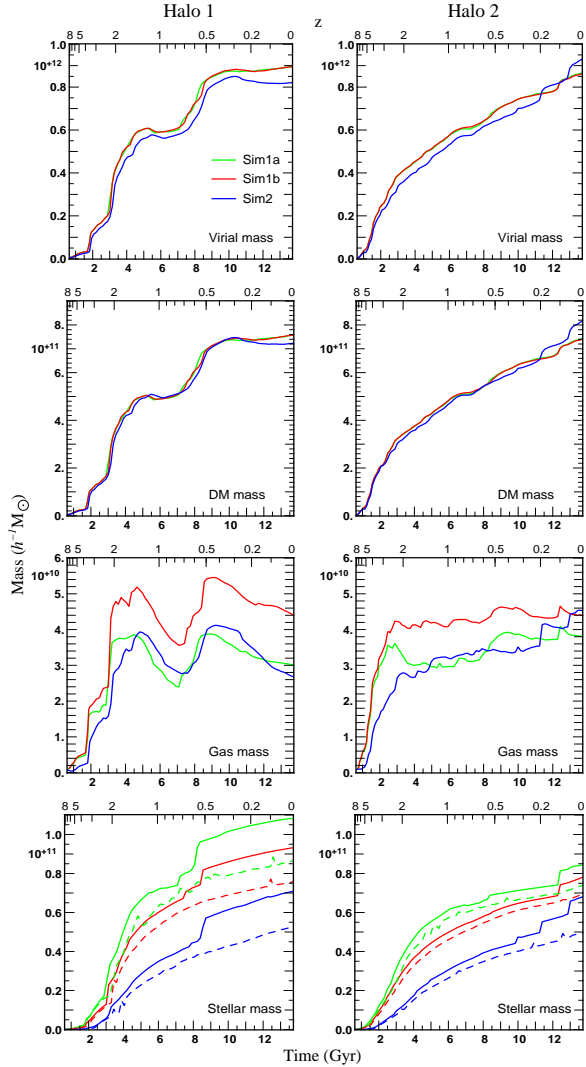}}
\caption{The evolution of the virial dark matter (DM), gas and stellar mass for \Halo1 (first column)
 and \Halo2 (second column) as a function of the time or redshift
from  $\Sim1a$ (green curves), $\Sim1b$ (red curves) and $\Sim2$ (blue curves).
{ In the stellar mass panels (fourth line) we also show in dashed lines the evolution of the stellar mass 
 inside the galaxy radius defined as $R_{gal}=0.1\,R_V$.}
Note that $Halo1$ undergoes several major merger episodes while $Halo2$ grows through 
smooth accretion. Moreover,
different feedback process prescriptions lead to different
baryon mass contents inside $R_v$.}
\label{mass}
\end{center}
 \end{figure}

The time evolution of the virial,  dark matter (DM), gas  and stellar masses
of each object at the virial radius and for the three simulations is shown in Figure\,\ref{mass}.
First, we notice that there is no particular difference between the evolution of
dark matter halo masses (and hence viral masses) between $\Sim1a$ and $\Sim1b$. However, the baryonic 
compositions are quite different. Indeed, stellar masses are significantly reduced in $\Sim1b$ 
due to higher amounts of energy released by the SN. The mass of the gas component is therefore higher in $\Sim1b$,
as expected.
For $\Sim2$, the stellar masses are even more reduced. This is mainly due to the fact that a significant fraction
of the gas has been expelled at high redshifts.

\begin{figure}
\begin{center}
\rotatebox{0}{\includegraphics[width=\columnwidth]{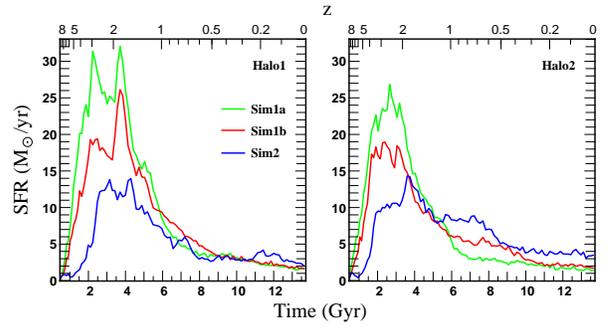}}
\caption{The evolution of the star formation rate (SFR) for $\Halo1$ (left panel) and $\Halo2$ 
(right panel) as labelled. 
}
\label{fig_sfr}
\end{center}
 \end{figure}

 Indeed, 
while the DM masses are quite similar between the 3 runs, the virial
 masses for objects in $\Sim2$ tend to be lower at high redshift, which indicate that a fraction of gas expelled at high z
does not recollapse at some  later stage. Interestingly, $\Halo2$ undergoes a minor merger
event at $z\simeq 0.1$, which corresponds for instance to a sudden variation in the evolution of its stellar mass.
This same accretion event
takes place at $z\simeq 0.2$ in $\Sim2$.
 Moreover, we can also identify clearly in $\Sim2$ another minor merger event at $z\simeq 0.05$
that has not occurred (yet?) in $\Sim1a$ and $\Sim1b$. This explains why the  dark matter mass (and therefore the
virial mass) of $\Halo2$ is higher in $\Sim2$
 than $\Sim1a$ or $\Sim1b$.
 This also suggests that small perturbations in the properties  of haloes at earlier times
can have some significant impact on the whole accretion history (see for instance, Thiebaut et al. 2008).            
It is also worth mentioning that although the two main central objects
have the same virial mass at $z=0$, they have undergone  different mass accretion
histories. Indeed, while $\Halo2$ grows regularly through smooth mass accretion,
the evolution of the mass of $\Halo1$ is more sporadic thanks to several major merger
events. It is therefore instructive to see to which extent these two different mass evolutions
could affect the evolution of the distribution of baryons in the vicinity of these two objects.

It is also worth studying how the stellar masses found in the different 
simulations  are related to the mass of their host dark matter haloes. From Table \ref{table2},
the mass of dark matter haloes varies from $1.02\times 10^{12} M_\odot$ to  $1.09\times 10^{12} M_\odot$
while according to observational and numerical analysis, the expected stellar mass range inside those
haloes is $M_* \simeq [2.5-3.5]\times10^{10}  M_\odot$ (Guo et al. 2010; Moster et al. 2010; 
Brook et al. 2012b).
In order to make a suitable comparison, instead of considering the stellar mass inside the virial 
radius $R_v$, we
define the galaxy radius as $R_{gal}=0.1\, R_v$ and all stars inside this radius contribute to
the galaxy stellar mass $M_*$. By doing this, we only select stars which belong to the central main galaxy 
 while excluding stars of accreted satellites inside $R_v$.  
  In Figure\,\ref{mass}, the evolution of each stellar mass inside  
$R_{gal}$ is shown in dashed line. One can see that at $z=0$, galaxy stellar mass ranges
are  $M_{*,\Sim1a} \simeq [1.1-1.2]\times10^{11}  M_\odot$,  $M_{*,\Sim1b} \simeq [9.9-1.1]\times10^{11}  M_\odot$ 
and  $M_{*,\Sim2} \simeq [7.0-7.4]\times10^{11}  M_\odot$ for $\Sim1a$, $\Sim1b$ and  $\Sim2$ respectively.
Thus,   each simulation tends to produce too many stars in the central region of haloes
and the trend is more severe in  $\Sim1a$ and  $\Sim1b$ relative to $\Sim2$.
This problem was already pointed out from other hydrodynamical simulations
which indicate that low feedbacks may result in more than an order-of-magnitude too many stars
(see for instance Piontek \& Steinmetz 2011; Brook et al. 2012b).
As a first conclusion, stronger and earlier feedbacks seem to be required in order to improve the stellar mass-halo mass
relation (also recently suggested by Stinson et al. 2012) and, in the present study, one probably also needs
 the action of additional sources of feedback in the late phase
of galaxy formation such as AGN to get even closer to the expected galaxy stellar masses.
Moreover the fact that our model 
leads to too high central stellar masses  suggests that the effect of all different sources
of feedback has been underestimated. Thus in the following,  baryonic fraction values found in our simulations may be 
slightly overestimated.


Finally, Figure\,\ref{fig_sfr} shows the evolution of the star formation rate (SFR) for 
each object and simulation. At $z=0$, SFR values are around $[1-5]$ $M_\odot/$yr  which is consistent with
observational values derived by Robitaille \& Whitney (2010) using {\sc Spitzer} data.
Note that these cosmic evolutions are similar for
the two haloes: at high redshifts ($z \geq 2$) 
the lower the SN efficiency  the higher the stellar mass. And the higher the
mass of formed stars  at high redshift, the higher the subsequent SN feedback.
This explains why, at low redshifts, a higher fraction of stars is produced in $\Sim2$ relative to
  $\Sim1a$ and $\Sim1b$.

\section{The baryon fraction in the LG}

\subsection{Spatial distribution in the simulated LG}

\begin{figure*}
\begin{center}
\rotatebox{0}{\includegraphics[width=2\columnwidth]{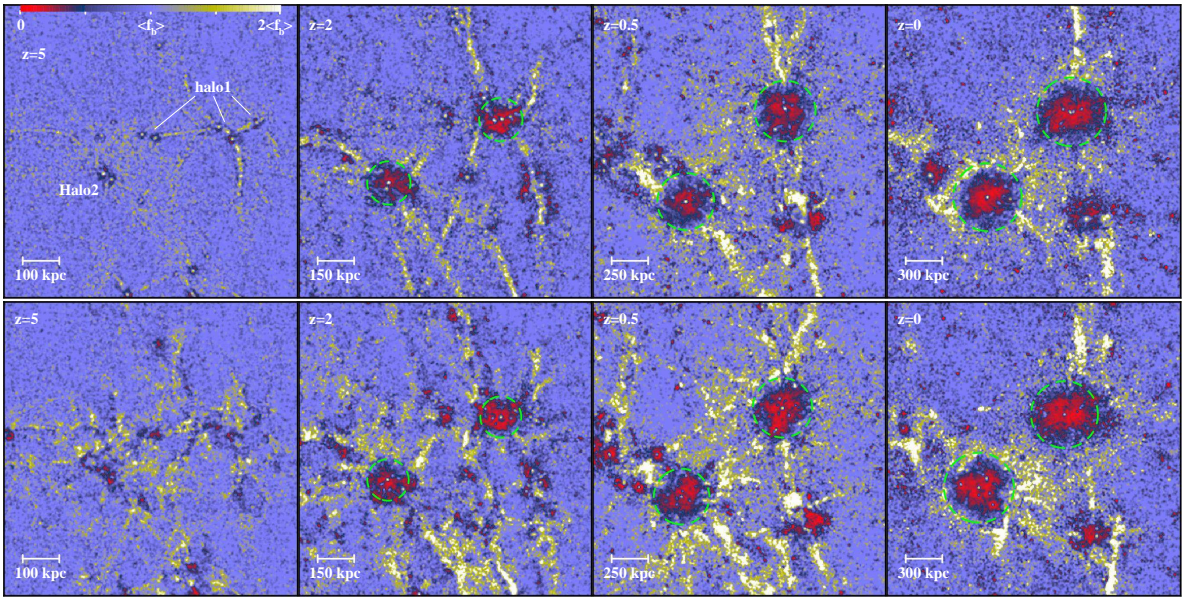}}
\caption{The projected baryonic fraction at $z=5$ (first column) $z=2$ (second column),
 $z=0.5$ (third column) and $z=0$ (fourth column)  from $\Sim1b$ (first line) and
 $\Sim2$  (second line). Dashed circles show virial radii. 
High $f_b$ value regions are clearly visible around galaxies (or proto-galaxies) and in
the filaments. This trend seems to be more pronounced in  $Sim2$ although
the difference between the two simulations is vanishing at low redshift.} 
\label{distribution1}
\end{center}
 \end{figure*}

 Fig.\,\ref{distribution1} shows the projected distribution of baryons in our LG type universes  
 derived from $\Sim1b$ and $\Sim2$ and at four specific epochs (i.e. $z=5, 2, 0.5$ and $0$).
 We clearly see  that the evolution of the distributions of baryons and dark matter
do not follow the same trend. Indeed, the regions of the universe in red and dark blue  
correspond to regions where the baryonic fraction is lower than the universal value. And
those specific regions are mainly located  around the forming protogalaxies or galaxies at each redshift,
while in the very inner part of dark matter haloes, $f_b$ is higher. This is an expected result:
due to cooling, the gas collapses to the center of haloes where stars can be  formed.
However, note that the size of these
 ``red'' cavities increases over time which suggests that
the fraction of gas that has collapsed to the center of the halo is not immediately replaced by some 
fresh gas from its vicinity.

Note also,  at high and low
redshift,  the existence of relative high baryon content regions which are located either in the filaments or
beyond the virial radii.  Such anisotropic distribution seems to be more pronounced in $\Sim2$ relative to $\Sim1b$. 
If the high baryon fractions in filaments can be understood by the dissipative nature
of gas, allowing it to cool to the dense filaments,  
it is crucial to characterise the mechanisms that drive  high baryon content regions
beyond the virial radii.

\begin{figure*}
\begin{center}
\rotatebox{0}{\includegraphics[width=0.9\columnwidth]{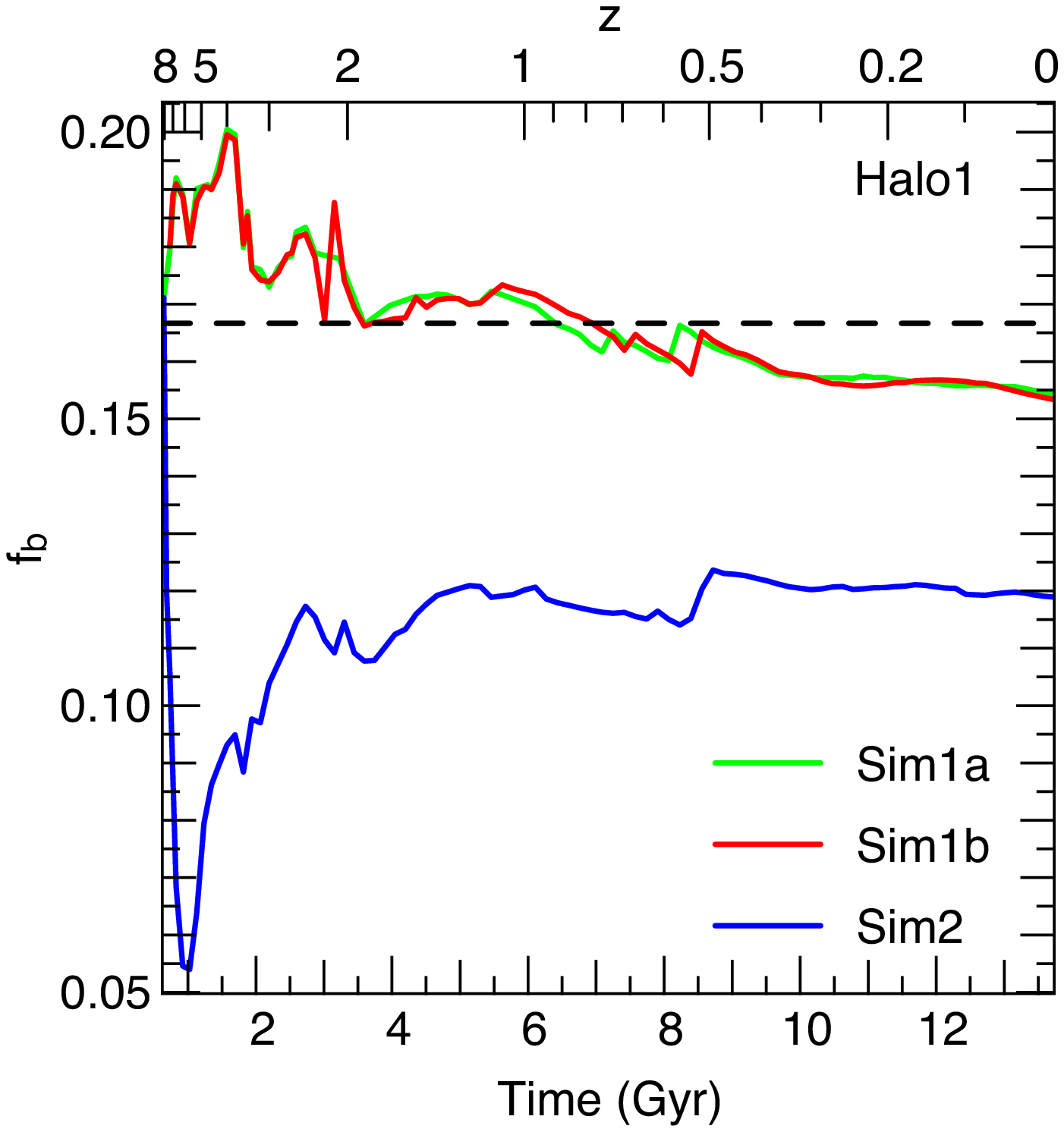}}
\hskip 0.5cm
\rotatebox{0}{\includegraphics[width=0.9\columnwidth]{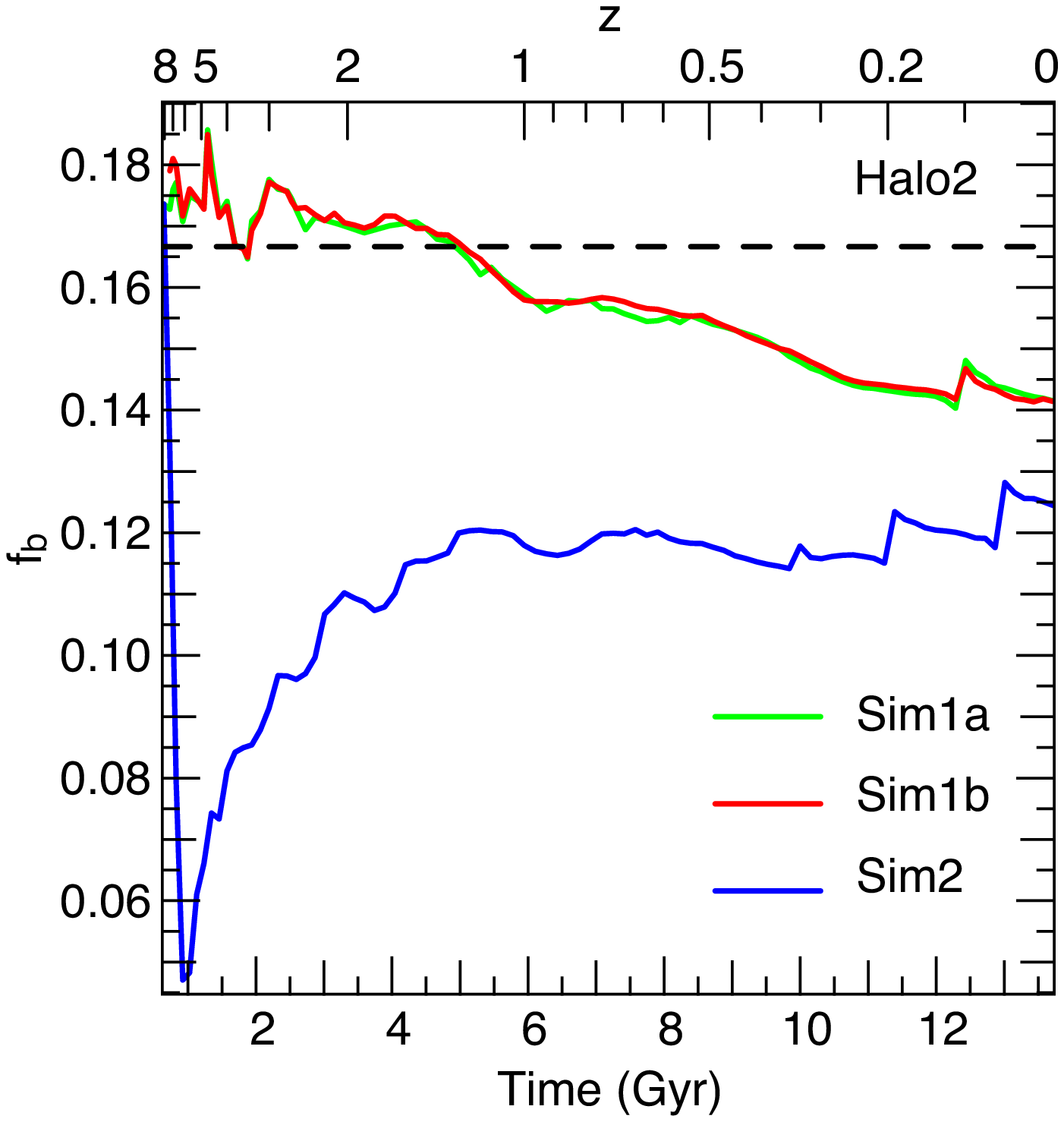}}
\caption{The evolution of  the baryonic fraction $f_b$ estimated at the virial radius for $\Halo1$ (left panel)
and $\Halo2$ (right panel) as a function of cosmic time. In each panel, red, green and blue lines correspond to values derived from
$\Sim1a$, $\Sim1b$ and $\Sim2$ respectively. The horizontal dashed line corresponds to the universal fraction.
Note that the early feedback model induces a lower ($f_b\sim 0.12$) baryon fraction at present time.}
\label{fb_evolution}
\end{center}
 \end{figure*}

\subsection{Cosmic evolution of $f_b$}

\subsubsection{The effects of feedback at low redshift}

 Let us first only consider the role of supernovae feedback in the evolution of the mass budget of Milky Way-type galaxies.
We therefore focus on both $\Sim1a$ and $\Sim1b$ in this sub-section.

The cosmic evolution of the baryonic fraction $f_b$ estimated at the virial radius for the two main haloes is shown
in Figure\,\ref{fb_evolution}.  In each of these two simulations, they follow the same trend. 
At high redshift, $f_b$ 
is close to the universal value and sometimes is slightly
higher. Cold flows provides gas to form stars. 
Note that the virial radius is estimated
according  to the local density, which is dominated by the newly formed stars. Since the resolution in these simulations
is limited, especially  for low mass systems, the virial radius may be underestimated,
 and subsequently this baryonic fraction may be in fact overestimated at high redshifts.

 From $z\sim 3$, $f_b$  is decreasing with cosmic time until it reaches the values of $\sim 0.15$
and $\sim 0.14$ at $z=0$ for $\Halo1$ and $\Halo2$ respectively.
These latter values are quite close to the universal value which is not surprising since
similar trends have  been already found from other hydrodynamical simulations using weak or no supernovae feedback,  for 
objects with a mass of $10^{12}\,h^{-1}M_\odot$ at $z=0$ (Crain et al. 2007;
Faucher-Gigu{\`e}re, Kere{\v s} \& Ma 2011). 
More interestingly,
no particular differences are seen in the evolution
of $f_b$ between $\Sim1a$ and $\Sim1b$.
 This suggests that higher SN feedback
 can reduce the star formation rate (and therefore the final stellar
mass) but is rather inefficient in expelling
the gas outside the virial radius at high and low redshifts for massive galaxies.
This results are in agreement with Stinson et al. (2011) who show that increased feedback
in haloes of such mass scale affects the star formation more than baryon content in the circumgalactic
medium (CGM). Moreover,  they show that higher feedback models can account for sufficient 
OVI and HI gas in the CGM compatible with the observed distributions. 
Therefore, in the present scenario,
 the relatively small differences in the baryon fraction between $Sim1a$ and $Sim1b$
could make  supernovae feedback prescriptions difficult to distinguish  and one
probably needs to focus on stellar mass and/or properties of the intergalactic medium
 in order to have better constraints (see also Dav{\'e}, Oppenheimer \& Sivanandam 2008; Shen, Wadsley 
\& Stinson 2010; Stinson et al. 2011,2012; Scannapieco et al. 2012; Brook et al. 2012b). In particular, it seems
that higher feedback of $Sim1b$ compared to $Sim1a$  may be necessary in reproducing values
closer to the correct stellar masses and properties of intergalactic medium as we will see in section 3.3.

\begin{figure}
\begin{center}
\rotatebox{0}{\includegraphics[width=\columnwidth]{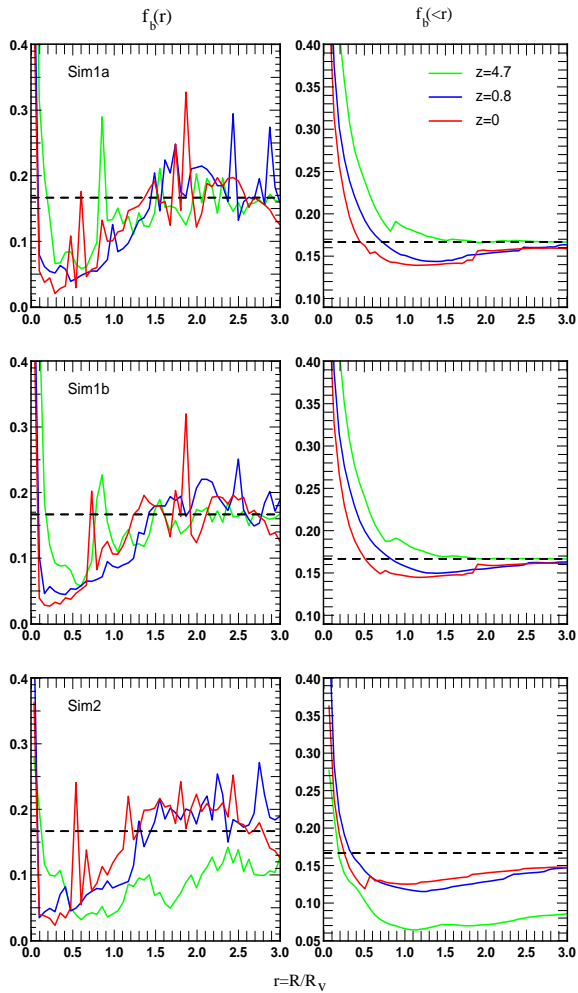}}
\caption{Distributions of the differential baryonic fraction, $f_b(R)$ (first column), 
and the integral baryonic fraction $f_b(<R)$ (second column) 
 with respect to the radius, $R$, for $\Halo2$ and  for three different redshifts, namely
$z=4.7$ (green lines), $z=0.8$ (blues lines) and $z=0$ (red lines).}
\label{fb_dis}
\end{center}
 \end{figure}

We have also studied the distribution of the integral 
 $f_b(<R)$ and differential $f_b(R)$ as respect to the radius $R$ at three different redshifts
$z=4.7$, $0.8$ and $0$ in Figure\,\ref{fb_dis}.
 Only results from $\Halo2$ are shown since similar trends are obtained for $\Halo1$.
At the center of each halo, $f_b$ is high due to the presence of the galaxy. 
At larger radii, $R\gtrsim 3R_v$, $f_b(<R)$ tends towards the universal value as expected. 
 Interestingly,  for high redshifts ($z>4.7$),  $f_b(<R)$ converges to the universal
value from above,
while this is not the case at lower redshifts, which suggest,  as mentioned above,
 that the gas infall is not recurrently replaced by some 
fresh gas from the outskirt of the haloes. Also, an excess of baryons at $(1.5-2.5)R_v$ can be clearly seen  in the
variations of the differential $f_b(R)$ for  different cosmic times, which correspond to 
regions in yellow  in Fig.\,\ref{distribution1}  around haloes. 
Similar plots were recently derived 
from observations using Chandra and Suzaku facilities for an isolated elliptical galaxy
with a $\sim$ Milky way mass (Humphrey et al. 2011) and a fossil group (Humphrey et al. 2012).

\subsubsection{The effects of feedback at high redshift}

In $\Sim2$ however, due to earlier and important eviction of 
gas from other potential sources of feedback (AGN, hypernovae,...) 
$f_b$ reaches its lowest value $f_b\sim 0.05$ at $z\sim 6$, 
then increases until $z\sim 1.5$ and becomes nearly  constant ($f_b\sim 0.12$)  until  the present time.  
This strongly suggests that sources of feedback acting at high redshift, even for a very short period,
can have a stronger impact on the final mass budget of  massive galaxies
than those acting in the later and longer phases of galaxy formation.
One can also notice from Figure\,\ref{fig_sfr}  that the gas outflow taking place at high redshift 
lead to the quenching of the star formation during a few million years ($5\leq z \leq 8 $).
This is in nice agreement with conclusions of Valiante, Schneider \& Maiolino (2012)
 using semi-analytical models of the formation and
evolution of high-redshift quasars. Indeed, according to their model, 
the observed outflow of a distant quasar at $z=6.4$ is dominated by quasar feedback
 which  {\it ``has considerably depleted 
the gas content of the host galaxy, leading to a down-turn in the star formation rate
at $z<7-8$''}.
Nevertheless, future observational proofs of star formation
quenching at these redshifts are required in order to confirm such statements.



\subsubsection{The role of  accretion of matter}

The time evolution of the baryonic fraction inside the virial radius depends on the competition
between the evolution of the mass that arrives within $R_v$, and the mass that leaves $R_v$. It raises the following question: are the low 
values of $f_b$  found at $z=0$ due to the fact that more baryons have been expelled out of the virial radius or
because more dark matter has been accreted inside $R_v$? To answer this question,
instead of studying the evolution of the mass
that enter or leave the virial radius at each subsequent step, we
focused on the location at a given redshift $z$ of all particles that have been accreted up to  this specific redshift.
To do this, we have first identified all the particles that have been accreted 
at each snapshot of the simulation from $z\sim8$ to ``$z$''. Then, we have computed the fraction of those particles
that are inside or outside $R_v$ at the considered redshift.
 By doing this, the interpretation of our findings will be easier since
 each accreted particle contributes only once in the final result. Indeed,  certain particles can be accreted,
 ejected and accreted again during  the whole evolution of these haloes.

\begin{figure}
\begin{center}
\rotatebox{0}{\includegraphics[width=1.\columnwidth]{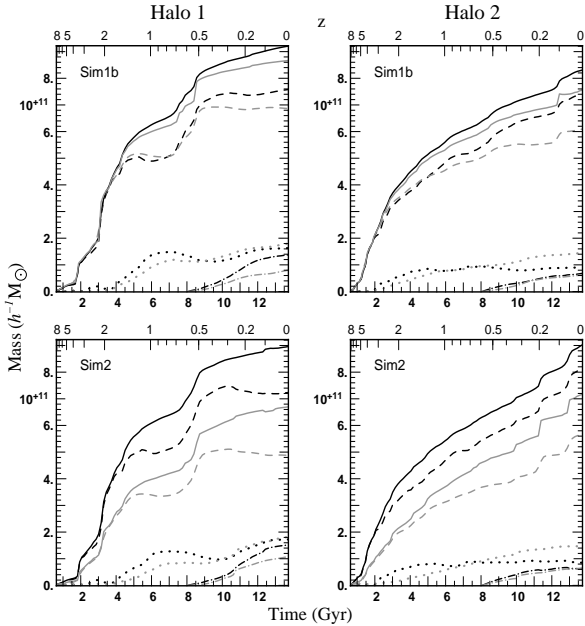}}
\caption{The evolutions of the total dark matter mass (solid black line) and  baryon mass (solid grey line) that was accreted
from $z\sim8$ to   a given redshift $z$ (see the main text for more details)
 for $\Halo1$ (first column) and $\Halo2$ (column 2),
derived from $\Sim1b$ (first line) and $\Sim2$ (second line). In each panel,
 the corresponding fractions of mass of DM (black color) and baryons (grey color) 
 that are inside $R_v$ (dashed lines) or outside $R_v$ (dotted line) is represented.
The dotted-dashed black and grey lines represent respectively the fraction of DM and baryons outside  $R_v$, corresponding to 
 particles accreted from $z\sim0.6$ to   a given redshift $z$ .
Note that the  baryon mass evolutions have been multiplied by $(\Omega_{\rm M}-\Omega_b)/\Omega_b$ to
facilitate a direct comparison with DM mass evolutions. }
\label{mass_acc2}
\end{center}
 \end{figure}

\begin{figure}
\begin{center}
\rotatebox{0}{\includegraphics[width=\columnwidth]{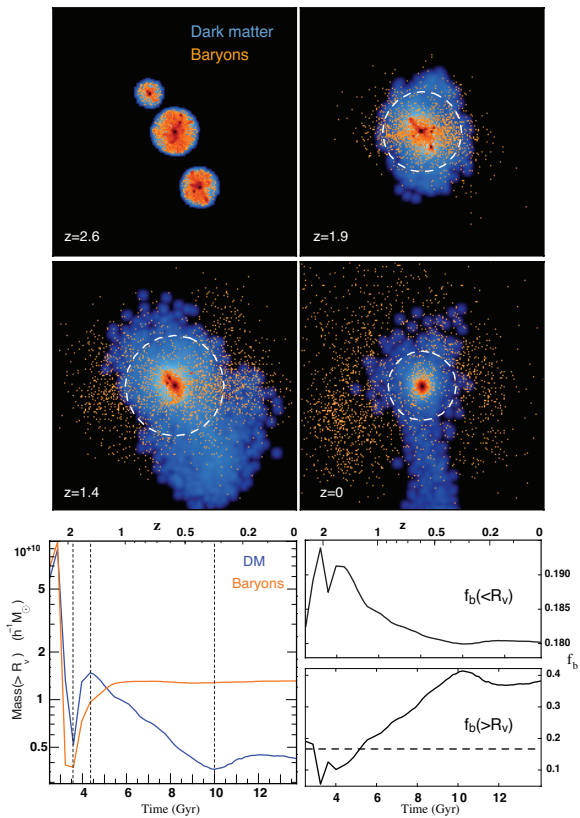}}
\caption{An illustration of the evolution of the distribution of baryons and dark matter
during and after given merger events. Panels
at $z=0$ is $1\times1$ Mpc across  while the other ones are $0.3\times0.3$
Mpc across. The white dashed lines represent the virial radii.
In the lower panels, we show the evolution of the mass of baryons
and DM outside the virial radius of the host halo
(left panel) and the evolution of the baryonic fraction
estimated from these 3 haloes  inside and outside the virial radius
(right panel).
One can see that a non negligible number of DM particles are re-accreted inside $R_v$ (from
$t\sim 4.5$ to $t\sim 10$ Gyr) which is not the case for baryons. As a result,
 the final values of  $f_b(<R_v)$ and $f_b(>R_v)$ 
are respectively lower and higher than initial ones.}
\label{illustration}
\end{center}
 \end{figure}

Figure\,\ref{mass_acc2} shows the evolution of the total accreted DM and baryon mass  from $z\sim8$ to   
a given redshift, $z$, for both haloes, and for both  $\Sim1b$ and $\Sim2$ (results derived from  $\Sim1a$ are not shown since they are similar to those from 
 $\Sim1b$). We also plot the fraction of these accreted particles that are located inside or outside $R_v$
 at this given redshift $z$.
First, we look at the evolution relative to $\Halo1$. One can see that a higher fraction of dark matter mass 
has been accreted from $z=8$ to $z=0$
relative to the baryon component (solid lines). Moreover, a similar fraction of dark matter and baryons are located outside the virial radius at $z=0$. That means that
the fraction of baryons released during the formation of $\Halo1$ is not higher than that of dark matter, which demonstrate that
the low value of $f_b$ at $z=0$  follows from the fact that more dark matter particles has been accreted over cosmic time.
This can be understood given the difference in nature between the dissipative gas and the non-dissipative dark matter. In particular, 
 the gas component can be shock-heated at the virial
radius which  considerably affects and slows down its further accretion on to the halo
 as already shown in hydrodynamical simulations using no feedback  (see for instance O{\~n}orbe et al. 2007). 
 In contrast, especially after an important episode of accretion or merger event, 
the dark matter component can ``damp out'' its orbital motion until it completely relaxes.
For $\Halo2$  the same trends are observed but the mass of baryons outside the virial radius 
turns out to be higher that that of dark matter. However
this does not necessarily mean that a higher fraction of baryons has been expelled through physical
mechanisms such as feedback or merger events. Indeed, if we represent the fraction of baryons and DM released
from $z\sim0.6$ to $z=0$ (see Figure\,\ref{mass_acc2}), one can see that same amounts of mass 
have been expelled at $z=0$ by the two components. 
It implies in fact that a significant fraction of dark matter  which has been expelled between $z=8$ and $z\sim0.6$
has been re-accreted between $z\sim0.6$ and $z=0$.  This is not the case
for baryons which are generally heated through the expelling process and therefore remain outside the virial radius afterwards.
In order to illustrate this, 
 Figure\,\ref{illustration} presents the evolution of the distribution of baryons and dark matter
for  \Halo1. From $z=2.6$, the main progenitor undergoes two mergers with mass ratios of
1:1.8 and 1:5.
Here,  we have selected all
particles inside the virial radius of each of the 3 objects. 
Note that the orbits of the two mergers are quite radial.
After the two merger events,   the spatial distributions of baryons and
dark matter are quite different. Most of the dark matter particles are located inside the virial radius,
though a small part has been expelled in the same orbital direction. The baryons tends to
either collapse toward the center or are expelled all around. 
This phenomenon is observed at $z=0$ where a non-negligible part of baryons 
have been expelled and remain outside the virial radius.
Figure\,\ref{illustration} also shows the evolution of the mass 
of DM and baryons located outside the virial radius. In the case of the
dark matter component, there are 3 distinct phases. First,  from the beginning of the experiment 
until $t=3$ Gyr, both satellites are accreted inside $R_v$. Then in the second phase (from $t=3$ to $t\sim4.5$ Gyr)
 some of the dark matter is expelled  via tidal disruption. Finally from $t\sim4.5$ Gyr
some of the dark matter particles are re-accreted. 
Note that evolution from $t\sim 10$ Gyr is probably due to the effect of another merger event.
However, the baryon component follows the two first phases but not the third one: once baryons are expelled, most of them
remain outside the virial radius.
In this specific example, we found that 
2.1\% and 5.8\% of dark matter and baryon masses respectively are released at $z=0$ from the three main objects.
This is in good agreement with results obtained from idealized simulations 
(Sinha \& Holley-Bockelmann 2009).
 It is also interesting to see that due to mergers, unbound gas diffuses far away from the galaxy; 
thus mergers  represent
 a potential mechanism for enriching the IGM with metals.
 Note that this enrichment must occur early, as in our model, to avoid disruption
 of the cold and weakly enriched Lyman alpha forest.
It is also worth mentioning that during a merger event, gas can be expelled via both tidal disruption and
outflows from feedback due to the merger driven starburst. However,  since there is no significant 
difference in the evolution of $f_b$ within the virial radius between $Sim1a$ and $Sim1b$, this
suggests that supernovae feedbacks have a lower contribution relative to tidal disruption for Milky Way
mass galaxies. As observed in the evolution of the dark matter component, tidal disruption should guarantee escape
whereas a starburst driven outflow is much more problematic (see for instance Powell, Slyz and Devriendt 2011).

Finally, it is interesting to point out that $\Halo1$ undergoes major merger events between  $z\sim0.8$ and $z\sim0.5$, leading
to important mass outflows from $z\sim0.5$ and $z=0$. In particular, Figure\,\ref{mass_acc2} indicates that more dark matter particles are released
than baryons after the merger events.
 Most of this dark matter is not yet re-accreted,  which explains why $\Halo1$ tends to have values of $f_b$  higher 
than $\Halo2$ at $z=0$. This also explains why in Figure\,\ref{distribution1} the high $f_b$ value regions around $\Halo1$ seem to be less
pronounced in comparison to those around $\Halo2$.

Thus,  the evolution of the mean baryonic fraction value $f_b$  at the virial radius
 is essentially governed by the relative efficiency at which the dark matter and baryons are accreted.
Indeed, the evolution of  the accreted mass of both dark matter and baryons follows two different regimes:
a rapid growth at high redshifts ($z\geq 1.5$)  and a slower one at lower redshifts. These evolutions are directly related
to the total mass evolutions in  Figure\,\ref{mass_acc2} which also presents these two regimes. 
This behaviour compares well to  the evolution of
individual dark matter haloes (see for instance Wechsler et al. 2002;  Vitvitska et al. 2002,
Peirani, Mohayaee \& Pacheco 2004,
McBride et al. 2009 and Tillson, Miller \& Devriendt 2011),
indicating a fast growth of the mass
for $z > 1.5 - 2.0$, followed by a phase where the accretion rate is slower. Hydrodynamical simulations
have also suggested that galaxy formation presents two phases:  a rapid
early phase ($z\geq 1.5$) during which stars mainly form within the galaxy from infalling cold gas
(Katz et al. 2003; Kere{\v s} et al. 2005;  Ocvirk, Pichon \& Teyssier 2008; Dekel et al. 2009) and a second one ($z\leq 1.5$) during which stars are accreted
(Oser et al. 2010; Hirschmann et al. 2012).
This latter phase is dominated by the accretion of diffuse matter and  small satellites (Fakhouri \& Ma 2010; 
Genel et al. 2010; Wang et al. 2011; van de Voort et al. 2011).
Hence the high $f_b$ values at high redshift derived from $\Sim1a$ and $\Sim1b$ can be explained by 
the accretion of high gas-to-dark matter ratio from dense region such as filament via cold flows.
But at lower redshift, $f_b$ is decreasing due to the accretion of low gas-to-dark matter ratio material, in 
particular from the diffuse region. Indeed, we found that the baryonic fraction
of the diffuse accreted matter (namely $f_b\sim0.11-0.14$) is on the average much lower than  the universal value:
 as the haloes become
more massive, the temperature at the virial radius increases and the gas is shock-heated and this process 
slows down its accretion onto the halo.  
For $\Sim2$,  $f_b$ reaches its lowest value ($f_b\sim 0.05$) at $z\sim5$ 
right after some significant expulsion of gas. After that, there is a short phase where  $f_b$ increases because
 the accreted matter has a higher baryonic content (but not necessarily higher than the universal value). 
Then,  in the late phase of evolution of each halo,    $f_b$  tends to be constant ($f_b\sim 0.12$)
 because the average accreted matter tends to have the same 
baryonic fraction value as the mean value inside  virial radii.

\subsection{Properties of the IGM}

\begin{figure}
\begin{center}
\rotatebox{0}{\includegraphics[width=0.9\columnwidth]{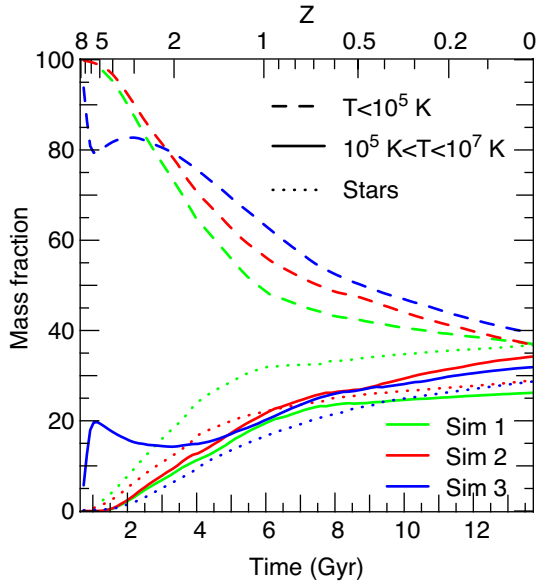}}
\caption{The evolution of the IGM components within a sphere  centred on the local Universe type haloes
and of comoving radius 1.5 Mpc/h: cold gas ($T<10^5\,K$, dashed lines),
 WHIM ($10^5\,K<T<10^7\,K$, solid lines) and stars (dotted lines). Green, red and
blue lines correspond to $\Sim1a$, $\Sim1b$ and $\Sim2$ respectively.
bf The evolution of  the hot gas ($T\geq 10^7$\,K) is not shown 
since this component represents less than $1\%$ of the total baryon mass budget
in the selected region.}
\label{whim}
\end{center}
 \end{figure}

\begin{figure}
\begin{center}
\rotatebox{0}{\includegraphics[width=\columnwidth]{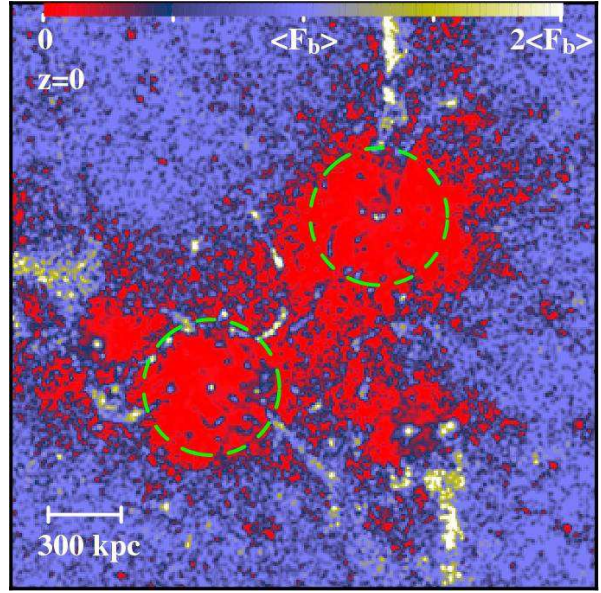}}
\caption{The projected baryonic fraction at $z=0$ derived from $\Sim1a$ when the WHIM is 
not taken into account in the estimation of $f_b$. }
\label{whim2}
\end{center}
 \end{figure}

The  WHIM has proven to be difficult to detect and represents a serious candidate
for the missing baryons. It is therefore instructive to 
estimate the fraction of the WHIM  in our  simulated local universes.
For this purpose, we have divided the Inter-Galactic Medium (IGM)
into four components: (1)  stars,
(2) cold gas ($T<10^5\,K$), (3) WHIM ($10^5\,K<T<10^7\,K$) and hot gas ($T>10^7\,K$).
Figure\,\ref{whim} shows the time evolution of three baryon components (stars, cold gas and WHIM)  within
a sphere  centred on the local Universe-type haloes and of comoving radius 1.5 Mpc/h.
 The evolution of  the hot gas ($T\geq 10^7$\,K) is not shown in Figure\,\ref{whim}
since this component represents less than $1\%$ of the total baryon mass budget
in our selected region, as expected since we are not in a cluster environment. 
It shows that the WHIM represents  $\sim 30$\% of the baryon budget at $z=0$, in good agreement 
with past work based on hydrodynamical simulations (Cen \& Ostriker 1999; Dav\'e et
al. 2001; Cen \& Ostriker 2006; Rasera \& Teyssier 2006; Tornatore et al. 2010) when AGN feedback is not taken into account.
We also found that the warm and hot gas becomes 
the dominant baryonic component at large radii ($R>150$ kpc) from each main halo,
in good agreement with previous observational analysis of massive galaxies
(Humphrey et al. 2011, 2012)  or with previous numerical simulations
 (van de Voort et al. 2011; van de Voort \& Schaye 2012).
However, when the contribution of the WHIM is not taken into account
 in the estimation of the baryonic fraction at the virial radius, namely when
 only the cold gas ($T\leq 10^5$\,K) and stars  are considered,
$f_b$ values obtained  at $z=0$  only decrease by $\sim 15$\%. This  follows from the 
fact that stellar masses obtained in our experiments have a large contribution
in the final $f_b$ values and thus we probably need the additional effect of AGN feedback 
to regulate the SFR and also increase the mass of WHIM inside the virial radius.  
Nevertheless,  the distribution of baryonic fraction seen in Figure\,\ref{distribution1} 
is significantly affected when the WHIM is not taken into account. Indeed,
Figure\,\ref{whim2} shows that in this latter case, the baryonic fraction is lower than
$0.5 \langle F_b
 \rangle $   in the direct vicinity of the virial radius. 

Outside the virial radii, our analysis in the previous section has suggested that
the IGM  is composed of gas that either
never collapses onto haloes or is released by mechanisms such as feedback  or merger events. 
But  what are the relative proportions? 
From Figure\,\ref{mass_acc2} 
we can estimate that $\gtrsim 20\%$ of the total accreted baryons is expelled
 from $z=8$ to $z=0$ for $\Halo1$ and $\Halo2$, which is quite significant.
These results are also in good agreement with Sinha \& Holley-Bockelmann (2010) who found from semi-analytic techniques 
that up to $\sim$ 25\% of gas can be unbind over the course of galaxy assembly.
From our analysis, we found that the gas that has been released after being accreted on to  haloes
is mainly located between $R_v$ and $2R_v$ at $z=0$. If we focus on $\Halo1$ which is quite isolated,
we found that the gas component between $R_v$ and $2R_v$  consists of 55.9\%,  58.6\% and  60.5\% 
of gas that has been expelled from
$\Halo1$ from $\Sim1a$, $\Sim1b$ and $\Sim2$ respectively. In other words  $\sim$ 40\% of this gas has never collapsed
onto $\Halo1$. It is also worth mentioning that $\gtrsim 95\%$ of the gas that has been released through merger events
or feedback is in the form 
of WHIM at $z=0$ but only $\sim$ 65\% in the case the gas that has never collapsed into the halo.
 Hence, the gas that has been ejected through different
mechanisms during the galaxy formation and evolution significantly increases the mass of the WHIM.

\section{Discussion and Conclusions}

In this paper, we have analysed  hydrodynamical simulations with an extensive
treatment of the physics of baryons in order to study the evolution of the
distribution of baryons in a local group-type object. 
In particular, we have investigated the impact
of i) supernova feedback,  ii) more powerful sources of feedback
at high redshift such as AGN from intermediate mass  black holes
or hypernovae and iii)  accretion/ejection and merger events
in the evolution of the  mass content of Milky-Way type galaxies.

First, when no high early feedback is considered as is the case in $Sim1a$ and $Sim1b$, we found that
 the evolution of the mean baryonic fraction value $f_b$ estimated at the virial radius
of the two main central objects is essentially governed by
 the relative efficiency at { which }  dark matter and baryons are accreted.
Indeed, in the first and early phase of galaxy formation (the cold mode of accretion $z\geq 1.5$) during which stars mainly form
within the galaxy from infalling cold gas (Katz et al. 2003; Kere{\v s} et al. 2005;
Ocvirk, Pichon \& Teyssier 2008; Dekel et al. 2009)
$f_b$ is in general higher than the universal value. But in the second phase (the hot mode of accretion), dominated by
the accretion of diffuse matter or small satellites, $f_b$ tends to decrease and 
is  10-15\% lower than the universal value, 0.166, at $z=0$. 
This decrease is not due to the fact that more baryons (relative to  DM) are expelled
beyond the virial radius by physical mechanisms such as feedback or merger events. Rather
this  essentially follows from the relative nature  of 
 dissipative gas and non-dissipative dark matter. Specifically, 
 the gas component can be shock-heated at the virial
radius, which  considerably slows down its further accretion into the halo (see for instance 
O{\~n}orbe et al. 2007). 
However, shock-heating will presumably not be a large factor in low mass galaxies,
but as mentioned in the introduction, feedbacks such as UV-background and supernovae winds
are supposed to play a major role in the effective reduction of the baryon content of these galaxies. 
Additionally, we found that accretion and particularly merger events can release
a fraction of both dark matter and baryons. But again, dark matter that has been expelled can be re-accreted
more efficiently  than the corresponding baryons, hereby enhancing
the decrease of $f_b$ inside the virial radius. 
Consequently, our simulations suggest that the gas located beyond the virial radius is mainly 
in the form of WHIM  that comes from either gas that never collapsed prior to the formation of 
the proto-galaxy ($\sim 40\%$)  or gas expelled by feedback and merger processes ($\sim 60\%$).

Our comparison of \Sim1a and \Sim1b also  suggest that  globally increasing somewhat the efficiency of supernovae 
 does not significantly  affect the evolution of $f_b$ inside the virial radius. In other words, standard SN feedback
is not powerful enough to expel the baryons out of the gravitational potential
 of dark matter haloes of such mass scales.
This also means that, if future observations  indicate that $f_b$ inside the virial radius of Milky Way mass galaxy
at $z=0$ is relatively close to the universal value, then
in order to make clear a distinction between the different feedback models, the stellar mass and
the  properties of the intergalactic medium may provide better
constraints than baryon fractions, as suggested by recent numerical works (Dav{\'e}, Oppenheimer \& Sivanandam 2008; 
Shen, Wadsley \& Stinson 2010; Stinson et al. 2011,2012; Scannapieco et al. 2012; Brook et al. 2012b).
 But  at high redshift  the situation is quite
different since haloes are less massive and thus feedback mechanisms  have
a more effective contribution. Indeed for our other scenario
(for which a higher fraction of gas is expelled at high redshift ($z\sim8$) by more powerful
 source of feedback 
such as  AGN from intermediate mass  black holes or hypernovae implemented in \Sim2) the accretion
 of gas is strongly affected at high redshift, and drives the
mean baryonic value inside the virial radius to a lower values at $z=0$. This is one important conclusion 
of this paper: to get a low  value of $f_b$ at present time, 
physical mechanisms able to expel the gas {\sl at high redshift } will 
have a stronger impact on the deficit of baryons in the mass budget 
of Milky Way type-galaxy today
 than those that expel the gas  in the longer,  late phases of galaxy formation.
Thus, if future observations favour a  
relatively low value of $f_b$ at $z=0$, then this ``high redshift feedback" scenario is the most probable.
 
However,  it this worth mentioning that
the effective
contribution of non-gravitational heating such as galactic winds
from supernovae in the evolution of $f_b$ within the virial
radius of Milky Way type galaxy is still a controversial and a debated topic in the recent literature.
From our modelling and experiments, 
we found that increasing the efficiency of supernova feedback leads
to the reduction of the stellar mass
but has no   particular impact on the evolution of $f_b(<R_v)$
 for system of such mass scales (cf Powell, Slyz \& Devriendt 2011). 
  Such statements were  also advocated by  Anderson \& Bregman  (2010) 
from observational constraints on the density of hot gas around the Milky Way.
Moreover, using cosmological simulations, Faucher-Gigu{\`e}re, Kere{\v s} \& Ma (2011)
have investigated the net baryonic accretion rates through $R_V$. Although
the net baryonic accretion rate is sensitive to galactic outflows, especially for low mass systems,
 the baryon mass fractions at $z=0$ for objects with a mass higher than $10^{12} M_\odot$
and for different SN feedback prescriptions have similar values to those derived in the present study. 
However, other hydrodynamical simulations from  Scannapieco et al. (2008, 2009) using higher SFR efficiency ($c_*=0.1$)
found that at $z=0$, the  baryon fractions inside the virial radius are $\leq 0.10$ within the virial radius
{\it ``indicating that a significant
amount of baryons has been lost through winds''}. 

AGN feedback is supposed to play an important role in the formation of massive objects
 such as giant ellipticals
 in galaxy groups (Tabor \& Binney
1993; Ciotti \& Ostriker 1997; Silk \& Rees 1998).  From previous numerical studies, the inclusion of AGN 
is expected to reduce the stellar mass  (see for instance Sijacki et al. 2007;
 Dubois et al. 2010, 2012; Hambrick et al. 2011b) and also 
increase the mass of WHIM (Cen \& Ostriker 2005; Tornatore et al. 2010; Durier \& Pacheco 2011;
 Roncarelli et al. 2012). 
Our study does not include the effect of AGN feedback at low redshift,
but we think that such sporadic events will not  have a major contribution 
in the baryonic fraction value inside  the virial radius at  present time.
Indeed, from observationally motivated constraints, AGN and SN at low to moderate redshift
seem not to produce the expected correlations with the baryonic Tully-Fisher relationship
 (Anderson \& Bregman 2010).
 Moreover, as mentioned above,
 observations of massive objects reported by Humphrey et al. (2011, 2012) 
 seem to suggest that AGN feedback has no particular impact on the baryonic fraction
values within the virial radius.
Also, recent hydrodynamical simulations show that AGN and SN feedback
may mutually weaken one another's effect by up to an order of magnitude in haloes in the mass range 
of $10^{11.25} - 10^{12.5} M_\odot$ (Booth \& Schaye 2012).
At high redshift, the effects of galactic winds on the IGM are also controversial.
In the present paper, our simple minded modelling indicates that galactic winds can expel some
gas outside the virial radius and affect significantly the evolution of the  baryon content
of massive galaxies. Similar conclusions have been recently drawn by Roncarelli et al. (2012)
from hydrodynamical simulations suggesting that galactic winds acting at earlier epochs can
prevent the IGM to collapse in dense structures. 
Also, as mentioned above, semi-analytical model from Valiante, Schneider \& Maiolino (2012)
indicates that the observed outflow of a distant quasar at $z=6.4$  (Maiolino  et al. 2012)
is expected to have considerably depleted the gas content of the host galaxy. 
However,
simulations conducted by Di Matteo et al. (2011) or Shen et al. (2012)
 found that galactic outflows 
at high redshift  may not be very effective at stopping the cold gas from penetrating
the central regions. 

In summary, 
our study indicates that in order to reach lower $f_b$ values at $z=0$ for  Milky Way type galaxies,
 the eviction of cold gas by feedbacks during the first phase of galaxy formation at high redshifts
 proves to be crucial. Our study also suggests that these  high redshift feedback mechanisms 
 more efficiently reduce  galaxy stellar mass in the central part of haloes and 
thus show better agreement to galaxy observable such as the stellar mass-halo mass relation.
 Moreover, efficient high redshift feedback processes  seem also to be 
required in order to slow down the growth of galaxies at high redshift
and thus to reproduce the observed number density evolution 
as recently pointed out by Weinmann et al. (2012). The presence of strong early feedbacks is    
also motivated in disk galaxy formation theory. Indeed, it has been suggested that early outflows could 
eject low angular momentum gas allowing disk galaxies to form (Binney, Gerhard and Silk (2001). 
Such hypothesis was confirmed by hydrodynamical simulations which show that outflows
can indeed remove preferentially low angular momentum material which can be re-accreted
latter with higher angular momentum and forms a disk (Brook et al. 2011, 2012a). 
If such statements are correct, 
numerical and observational efforts have to be focused   towards characterising the respective
role of each feedback process  on the IGM at high redshift.  In particular,
are the baryons expelled from  protogalaxies or do they never collapse into the gravitational
potential well of these objects?  In the present paper, we only focus on
the first mechanism,
 but it is not excluded that processes such as photoionization 
 play a major role especially at high redshifts  for low mass systems.
In this regard, it is interesting to compare our results with those of  Guedes et al. (2011)
who have studied the formation of late-type spirals from the ERIS simulation.  Specifically, they claimed
that the collapse of baryons is heavily suppressed at high redshift  by the UV background.
And it is particularly  instructive to see that from their highest mass resolution simulation,
 the evolution of the baryonic fraction (see their Figure 6 in  Guedes et al. 2011)
is very similar to those we obtained from $\Sim2$.  They also found a similar value at $z=0$ namely $f_b\sim0.7 <f_b>$.
Our simulations do include UV background but our mass resolution may be too small to account for this process 
properly. 

More detailed and well-resolved hydrodynamical simulations
are required to investigate the evolution of the mass content of galaxies  at high and low redshifts.
But in order to achieve this goal, one must first improve the constraints of the expected energy injection
to the IGM from SN, AGN, UV background (etc...) via  observations.

\vspace{1.0cm}

\noindent
{\bf Acknowledgements}

\noindent
I.\,J.   acknowledges support from the Global Internship program by the
National Research Foundation of Korea (NRF).
We warmly thank the referee for an insightful
review that considerably improved the quality of the original
 manuscript.
We warmly thank D.\,Le Borgne, J.\,Devriendt, Y.\,Dubois, R.\,Gavazzi, 
T.\,Kimm, G.\,Mamon, J.A.\,de Freitas Pacheco  and T.\,Sousbie for interesting discussions.
We also thank D.\,Munro for
freely distributing his Yorick programming language (available at
\texttt{http://yorick.sourceforge.net/}) which was used during the
course of this work. This work was carried within the framework of the
Horizon project (\texttt{http://www.projet-horizon.fr}).

\end{document}